\documentclass[14pt]{article}
\usepackage{arxiv}

\usepackage[utf8]{inputenc} 
\usepackage[T1]{fontenc}    
\usepackage{hyperref}       
\usepackage{url}            
\usepackage{booktabs}       
\usepackage{amsfonts}       
\usepackage{amsmath,amssymb,amsfonts}
\usepackage{nicefrac}       
\usepackage{microtype}    
\usepackage{graphicx}
\usepackage{algorithmic}
\usepackage{algorithm}
\usepackage{subfigure}
\usepackage{wrapfig}
\usepackage{xcolor}
\usepackage{xspace}

\def\eqref#1{equation~\ref{#1}}

\def\1{\bm{1}}

\newcommand{\ie}{\textit{i.e.,}\@\xspace}
\newcommand{\eg}{\textit{e.g.,}\@\xspace}
\newcommand{\etal}{\textit{et al.}\@\xspace}

\title{Generative Extraction of Audio Classifiers for Speaker Identification}

\date{} 

\author{Tejumade Afonja\thanks{Work done while an intern at the University of Toronto and Vector Institute.},~ Lucas Bourtoule\thanks{Work done while a graduate student at the University of Toronto and Vector Institute.},~ Varun Chandrasekaran\thanks{University of Wisconsin-Madison.},~ Sageev Oore\thanks{Dalhousie University and Vector Institute.},~ Nicolas Papernot\thanks{University of Toronto and Vector Institute.}}

\begin{document}
\maketitle

\begin{abstract}
It is perhaps no longer surprising that machine learning models, especially deep neural networks, are particularly vulnerable to attacks. One such vulnerability that has been well studied is model extraction: a phenomenon in which the attacker attempts to steal a victim's model by training a surrogate model to mimic the decision boundaries of the victim model. Previous works have demonstrated the effectiveness of such an attack and its devastating consequences, but much of this work has been done primarily for image and text processing tasks. Our work is the first attempt to perform model extraction on {\em audio classification models}. We are motivated by an attacker whose goal is to mimic the behavior of the victim's model trained to identify a speaker. This is particularly problematic in security-sensitive domains such as biometric authentication. We find that prior model extraction techniques, where the attacker \textit{naively} uses a proxy dataset to attack a potential victim's model, fail. We therefore propose the use of a generative model to create a sufficiently large and diverse pool of synthetic attack queries. We find that our approach is able to extract a victim's model trained on \texttt{LibriSpeech} using queries synthesized with a proxy dataset based off of \texttt{VoxCeleb}; we achieve a test accuracy of 84.41\% with a budget of 3 million queries.
\end{abstract}

\section{Introduction}
\label{sec:introduction}

Machine learning (ML) models are often trained on large volumes of sensitive or proprietary data, and incur significant costs in development to their owners. However, through the simple process of responding to inputs (via a query interface) with output vectors, models leak information; Tramer \etal~\cite{tramer2016stealing} propose {\em model extraction} as the process of reverse engineering the model's confidential parameters through interaction via these query interfaces. Once a malicious actor gains access to a model, they can launch additional attacks through white-box access~\cite{szegedy2013intriguing,biggio2013evasion}.

Model extraction was shown to be successful for simple hypothesis classes~\cite{tramer2016stealing,chandrasekaran2020exploring}. More recently, Jagielski \etal~\cite{jagielski2020high} demonstrate extraction for deep neural networks (DNNs), for which the incentive for extraction is stronger due to the larger cost of training. Prior work~\cite{hitaj2018stolen, orekondy2019knockoff,juuti2019prada} demonstrate impressive results on vision task. In our work, we explore whether these results hold for other domains. In particular, we ask the following question: \textit{can the model extraction attack also be effectively performed in the audio (wave) domain?} Note that this differs from asking whether we can apply vision-based extraction on spectral \ie time vs. frequency representations of audio. While valid, the latter approach could confound issues intrinsic to audio with those relating to the image-like representation; we discuss audio representations in Appendix~\ref{B}. Instead, analogously to vision systems operating on pixels, we focus on end-to-end audio systems that work with waveforms.

In the first part of our paper, we justify why launching model extraction attacks on waveform-based audio classification models are a particular challenge. We compared an audio model trained to identify a speaker with its image processing counterpart for image classification, and found that existing model extraction techniques developed for image recognition models (in the vision domain) are not directly applicable to facilitate extraction of speaker identification models (in the audio domain). We find that the success of these vision-based model extraction techniques is due to the {\em transferability of the learned features}. This phenomenon is amplified by the usage of specialized convolution-based architectures for vision tasks, which are known to encode implicit biases about how people process images. Since such specialized architectures have not yet been developed for audio, we resort to better understanding what features learned are by our audio model. 

Our results suggest that a subset of the learned features are task dependent, \ie deeper layers reconstruct the representation specifically for the frequency contour of the particular speaker (see \S \ref{sec:4.3}) . We note that this task dependency poses a significant challenge to the success of model extraction, particularly approaches that rely on proxy datasets to learn the features of a victim model $\mathcal{M}$ from proxy set $\mathcal{D'}$ (which may be from a different distribution). Aside from the architectural bias in the vision domain, the difficulty of extraction may also be exacerbated by the lack of a single large, publicly available proxy dataset \ie audio corpus, that is well labeled and covers a wide range of audio tasks such as speech, music, environmental sounds, etc.~\cite{ldc2021,campbell1995testing,gemmeke2017audio}; model extraction in the vision domain often relies on such data to make meaningful and informative queries to the victim model. 

Based on these findings, we explored the use of synthetic datasets to perform model extraction attack using an attack strategy similar to that used in~\cite{papernot2017practical}. The attacker first trains a generative model using a proxy dataset $\mathcal{D'}$, and then samples this model to generate a set of queries $\mathcal{D''}$. These are sent to the victim model $\mathcal{M}$ for labelling. Finally, the attacker trains a surrogate model based on the synthetically generated, labelled queries. We find that this approach works significantly better than the naive approach of using the proxy dataset in itself for labeling and subsequent extraction. Furthermore, since generative models allow us to generate a large number of synthetic queries from the fixed size proxy dataset, \ie |$\mathcal{D''}$| $\gg$ |$\mathcal{D'}$|, we expect that these synthetic queries trained on $\mathcal{D'}$ are more likely to provide the necessary {\em coverage} \ie obtain a more diverse set of labels. 

However, developing such powerful generative model is non-trivial for two reasons: 1. sampling from a generative model with high fidelity is prohibitively expensive, making it difficult to use state-of-the-art autoregressive models; and 2. computationally efficient but low-quality generators such as generative adversarial networks often suffer from poor coverage of the data distribution, \ie mode collapse. To address the second challenge, we propose and experiment with two sampling strategies presented in \S~\ref{sec:extraction:gan-synthesis}: we use all the synthetic queries for labeling, but are selective in the specific query-label samples we ultimately use for extraction. Using this approach, we have found that a large number of queries from a fast but generative model with low fidelity gives the best extraction results.

To summarize, our contributions are as follows:
\vspace{-0.05in}
\begin{itemize}
\itemsep0em
\item We compared features learned from an image classification model with those learned from a speaker identification model, and found that the latter extracts more task dependent features. Our results shed light on what features a speaker identification model learns, a relatively new question.\footnote{Interpretability has been studied extensively in the vision domain~\cite{zeiler2014visualizing,yosinski2015understanding,simonyan2013deep, olah2017feature}. This has helped understand the types of features (such as those related to spatial properties) learned by deep neural networks while processing images and why these features often transfer well from one dataset to another. However, while some work has explored and continues to explore questions of what audio classifiers learn from their inputs~(e.g.,~\cite{dieleman2014recommending,dieleman2014endtoend}), it has not been studied nearly as extensively as the analogous question for vision.}
\item We introduce model extraction for speaker identification, a first attempt at extracting a waveform-based speaker identification audio model. We show that naive application of extraction attacks proposed for the vision domain fail, and therefore propose an approach utilizing generative models for synthesizing queries. 
\item By carefully selecting the synthetically generated queries that we retain for training the surrogate model using our proposed threshold-based sampling strategy, we show that model extraction for waveform-based speaker identification models is possible with high accuracy.
\end{itemize}

\section{Related Work}
\label{sec:rw}

Most work studying the vulnerability of ML applied to audio has focused on adversarial examples~\cite{carlini2018audio,taori2019targeted, du2019sirenattack, yakura2018robust}, where perturbations added to an originally correctly classified audio sample induces incorrect predictions. While adversarial examples target the integrity of a model's prediction, model extraction instead focuses on the confidentiality of the model's parameters. Prior work on model extraction has focused on vision (as discussed in \S~\ref{sec:introduction}), while our work focuses on the audio domain. 

Our study in \S~\ref{sec:motivation} is motivated by prior analyses of features learned by audio models. Lee \etal~\cite{lee2009unsupervised} illustrated first-layer bases learned by a convolutional deep belief network on unlabeled spectral TIMIT data~\cite{garofolo1993timit} and found correspondences to phones/phonemes. Dieleman \etal~\cite{dieleman2014recommending} trained a convolutional neural network (CNN) on music classification using mel-spectrograms: they found songs that activated the learned features, and they visualized slices of mel-spectral inputs that low-level features were sensitive to, including individual pitch detectors and certain harmonic detectors (\eg combinations of pitches). Dieleman \etal~\cite{dieleman2014endtoend} also compared spectral and wave-based models and presented interesting observations including mel-scale-like frequency-selective features in the raw audio model.

Prior efforts on generative modeling of speech are relevant to our model extraction approach. Specifically, we focused on two prominent works; Donahue \etal~\cite{donahue2019adversarial} worked on translating popular CNN-based GAN architectures from vision to audio. In particular, they established a correspondence between 2D and 1D layers in order to directly generate audio waveforms. Their method alleviates the need to approximately recover the phase from the amplitude spectrogram using the Griffin-Lim algorithm. We employ their technique to generate synthetic audio. The work of van den Oord \etal~\cite{oord2016wavenet} models the waveform of the audio signal one sample at a time (autoregressively), and can generate speech waveforms with subjective naturalness using dilated causal convolutions to deal with the modeling of long-range temporal dependencies. The sampling cost of such models makes them unsuitable for our proposed model extraction attack.

\section{Problem Setup and Threat Model}
\label{sec:threat}

\paragraph{Problem setup.} We focus on {\em speaker identification}---the task of identifying a speaker based on their voice---as this poses a challenging extraction problem with security-critical implications. Since different audio tasks have differently distributed data (\eg urban sounds, musical genres, etc.), we do not claim our results are broadly applicable to models trained for other audio tasks.

As the first exploration of audio extraction, having chosen the task, and having chosen an input representation (\ie sampled waveforms are fundamental to raw audio as pixels are to images; more discussion on audio representations is included in Appendix~\ref{B}), we next choose architectures for extraction.

Unlike in vision, where there is a long history of studying effective domain-specific architectures (\eg stacks of convolutions), we do not yet fully understand what architectures are best suited for waveform-based audio processing (though impressive steps have been made in recent years in both generation and discrimination~\cite{purwins2019deep}, and we do have evidence that, \eg priors imposed by architecture have significant effect on audio processing~\cite{pons2019randomly}).
While some audio classification models~\cite{chen2019end, lee2017raw} have shown comparable and sometimes better results using raw audio waveforms, many current approaches for speaker identification~\cite{snyder2018xvectors} do not yet make use of waveform representations. 
Since our focus is on extraction, rather than developing a highly specialized waveform-based architecture for speaker identification task, we choose a simple yet effective end-to-end architecture for speaker classification~\cite{knagg_building_2018}.

\paragraph{Threat model.} Model extraction in the audio domain involves two actors: (1) the adversary whose goal is to steal a victim model, and (2) the victim/oracle that owns the model (to be stolen) which honestly respond to each query it receives. Similar to prior work in vision~\cite{jagielski2020high}, we make several assumptions about the adversary’s knowledge and capabilities. These assumptions allow us to model a \textit{worst-case} attack which yields a stricter evaluation of the vulnerability of models from a defender's perspective. This is also a good way to focus on the peculiarities of audio in a model extraction context which is the aim of this work. More concretely, we assume: 

\noindent({\bf A1}) The attacker knows the exact architecture of the oracle as well as the optimization algorithm (\eg SGD with Momentum) and hyperparameters (\eg the learning rate, the batch size, and the number of epochs) used to train it. This strong hypothesis is common in the literature~\cite{chandrasekaran2020exploring,jagielski2020high}.

\noindent({\bf A2}) The attacker has access to a proxy dataset that is similar in size and purpose to the dataset that was used to train the oracle. This places this work in between dataset-aware extraction and data-free extraction~\cite{truong2020data}. Varying degrees of knowledge of this proxy dataset may be considered and expressed in terms of (a) the volume (number of samples) available to the adversary, (b) the coverage of the data \ie the number of classes it activates for the oracle, and (c) the quality of this data. All three factors are essential, as each factor introduces complexity in learning from audio.

\noindent({\bf A3}) We consider the case where the adversary knows the ground truth distribution of labels in the dataset that was used to train the oracle model. However, this is not a strong requirement: we also show that extraction is still possible without such assumption of the ground truth knowledge.

We consider a learning-based adversary who uses the oracle to label data, which is then used to train a local copy of the victim model. To do so, the attacker samples data from the proxy dataset $\mathcal{D}'$, and uses it to query the oracle classifier. The oracle responds with probability vectors (\ie soft or hard labels). This creates a new, labeled dataset $\mathcal{D''}$. The attacker now uses $\mathcal{D}''$ to train a copy of the oracle in a supervised fashion. When using soft labels, the loss function must be adapted to use probabilities instead of plain labels. This process is known as soft-label distillation~\cite{DBLP:journals/corr/HintonVD15}. To evaluate performance of the extracted model, we compare its accuracy (on some held-out set) with that of the oracle model.

\section{Motivation}
\label{sec:motivation}

In this section, we discuss our insight that enables audio extraction.

\subsection{Naive Extraction Approaches}
\label{sec:motivation:naive}

We apply a representative set of approaches for model extraction---developed mostly for vision (and NLP) domains---to audio model extraction for speaker identification. This not only serves to build our intuition for the audio modality, but also as baselines for evaluation of our proposed approach (see \S~\ref{sec:eval}). 

Our victim model is a \texttt{KnaggCNN} trained on \texttt{LibriSpeech}. Queries come directly from a proxy dataset, \texttt{VoxCeleb}, or from applying augmentation techniques to this proxy dataset. Both the victim's training dataset and the proxy dataset are popular datasets used for ASI in prior work~\cite{ravanelli2019speaker}. Refer to Appendices~ \ref{data:librispeech} and~\ref{model:knagg} for more details about the datasets and architectures used in our work. 

The data augmentation techniques we consider are:

\begin{itemize}
\item[{B1.}] {\bf Random Amplification:} Each example is amplified independently, with probability $p$ and by a factor $b$, where $b \sim U([1-a,1+a])$ and $a\in[0,1]$ is a parameter. 
\item[{B2.}] {\bf Pitch Shifting:} Each example is pitch-shifted independently, with probability $p$ and by $s$ semitones, where $s \sim \mathcal{N}(\mu=0,\sigma=1)$. We use the \texttt{Librosa} library~\cite{mcfee2015librosa} to achieve this. 
\item[{B3.}] {\bf Interpolation:} Similarly to Zhang \etal~\cite{zhang2017mixup}, we generate samples by linearly interpolating between random pairs of examples in the dataset.
\item[{B4.}] {\bf Baseline:} Queries are sampled from Gaussian noise. This is motivated by Krishna \etal's work~\cite{krishna2019thieves} demonstrating extraction of NLP models with gibberish data queries, \eg non-grammatical sentences made of randomly sampled words.
\end{itemize}

\begin{table}[ht]
\centering
\begin{tabular}{lc}
\toprule
{\bf Technique} & {\bf Accuracy (\%)} \\
\midrule
\midrule
{\em Proxy data only (full dataset)} & \textbf{64.80} \\ 
Proxy data only (subset) & 38.78 \\ 
({B1.}) Random Amplification ($a=0.2$, $p=1$) & 69.02 \\
({B1.}) Random Amplification ($a=0.5$, $p=0.5$) & 65.21 \\
({B1.}) Random Amplification ($a=0.2$, $p=0.5$) & 68.12 \\
({B2.}) Gaussian Pitch Shift ($\sigma=1$, $p=1$) & 38.82 \\ 
({B2.}) Gaussian Pitch Shift ($\sigma=1$, $p=0.5$) & 59.61\\ ({B3.}) Interpolation & 0.00\\ 
({B4.}) Gaussian Noise & 0.00 \\ 
\bottomrule
\end{tabular}
\caption{Augmentation techniques in the audio domain.}
\label{tab:augmentation}
\end{table}

Our results are summarized in Table~\ref{tab:augmentation}. Observe that: 
\begin{enumerate}
\item Random amplification ({B1}) does bring a small accuracy increase. We conjecture this to be the case as amplification may enhance the utility of specific features. This is generally contrary to what had been observed in the vision domain~\cite{jagielski2020high}. 
\item Augmentation techniques ({B2, B3}) are detrimental to extraction. This may stem from how these techniques remove any semantic information associated with the inputs (\ie pronouncing the word "sound" vs. pronouncing a randomly modulated variant of the word). 
\item Extraction with random noise ({B4}) fails in comparison to extraction with proxy data, again because of the lack of relevant semantic information. When we heard the generated audio samples, it sounded like flowing water, and unlike the required human voice-like audio.
\end{enumerate}

In summary, while techniques from prior work do not directly port, carefully adjusted data augmentation strategies may still help increase extraction efficacy, but they do not allow the attacker to fully bridge the accuracy gap.

\subsection{Coverage, Volume, and Task Dependence}
\label{sec:task-dependence}

Results involving a subset of the proxy dataset (row 2 in Table~\ref{tab:augmentation}) teaches us a useful lesson: the number of data samples (\ie \textit{volume}) of data plays an important role. We also observe that the proxy data's efficacy in capturing the victim model's distribution captured by a metric known as \textit{coverage} \ie the ratio between the size of the set of {\em unique labels} predicted by the victim on $\mathcal{D'}$, denoted ${|\mathcal{M}(\mathcal{D'})|}$ and the number of unique labels in the original dataset $\mathcal{D}$, denoted ${|\mathcal{M}(\mathcal{D})|}$ is important; we measure this to understand our exploration of the sample space.

\noindent{\bf Setup:} To isolate these factors, we conduct the following experiments: 

\begin{enumerate}
\item[{\bf E1}.] To understand the importance of coverage, we compare model extraction efficacy using two different subsets of the victim's {\em own} training set. One subset is diverse and still retains all the speakers (\ie high coverage), while the other subset only retains 200 of these speakers (\ie low coverage). 
\item[{\bf E2}.] To understand the impact of volume, we contrast these results with the extraction of the same victim, but with proxy datasets of differing volume.
\end{enumerate}

\noindent{\bf Results:} The results are summarized in Table~\ref{tab:coverage}, 

\begin{enumerate}
\item[Result for {\bf E1}:] Extraction with the diverse subset performs significantly better than with the less diverse subset. When we compare these two subsets with a random subset of $\mathcal{D}'$ (\ie \texttt{VoxCeleb}) having the same size, we see that coverage and extraction test accuracy are correlated. High coverage proxy datasets are provide more information during the extraction process (due to diversity in input features and output labels), ergo validating their worth.

\item[Result for {\bf E2}:] It is, however, difficult to understand how volume influences coverage: our experiments with {\em real data} suggest that increasing volume often tends to increase coverage as well (compare the two lines with \texttt{VoxCeleb} in Table~\ref{tab:coverage}), whereas our results with synthetic data in \S~\ref{sec:eval} (see Table~\ref{tab:thresholding_result}) indicate no significant correlation between volume and coverage. To summarize: volume by itself does result in some increase in extraction accuracy, but larger gains are obtained when increase in volume leads to increase in coverage (which results in commensurate gains during extraction).
\end{enumerate}

\begin{table}[t]
\centering
\begin{tabular}{lccc} 
\toprule
{\bf Proxy Data} & {\bf Volume} & {\bf Coverage (\%)} & {\bf Extraction Test Accuracy (\%)} \\
\midrule
\midrule
({\bf E1.}) \texttt{LibriSpeech} (diverse subset) & 19,160 & 98.3 & 80.06 \\ 
({\bf E1.}) \texttt{LibriSpeech} (200 speakers) & 19,160 & 49.3 & 37.58 \\ 
({\bf E2.}) \texttt{VoxCeleb} (full dataset) & 138,361 & 73.2 & 64.80 \\ 
({\bf E2.}) \texttt{VoxCeleb} (subset) & 19,160 & 54.0 & 38.78 \\ 
\bottomrule
\end{tabular}
\caption{Influence of coverage on the extraction test accuracy (after 40 epochs of training).}
\label{tab:coverage}
\end{table}

This suggests that coverage is essential, and at least a subset of the features learned by audio models are specific to the classes (here speakers) found in their training dataset. This would explain why it is important for the proxy dataset to cover most of the speaker classes in order to effectively extract the victim model. Since it is difficult to estimate the influence of training points (from a given class) on individual features learned by the model~\cite{koh2017understanding}, we reason about this influence as a function of what is learnt at different layers.

\begin{figure}[ht]
\centering
\includegraphics[width=\textwidth]{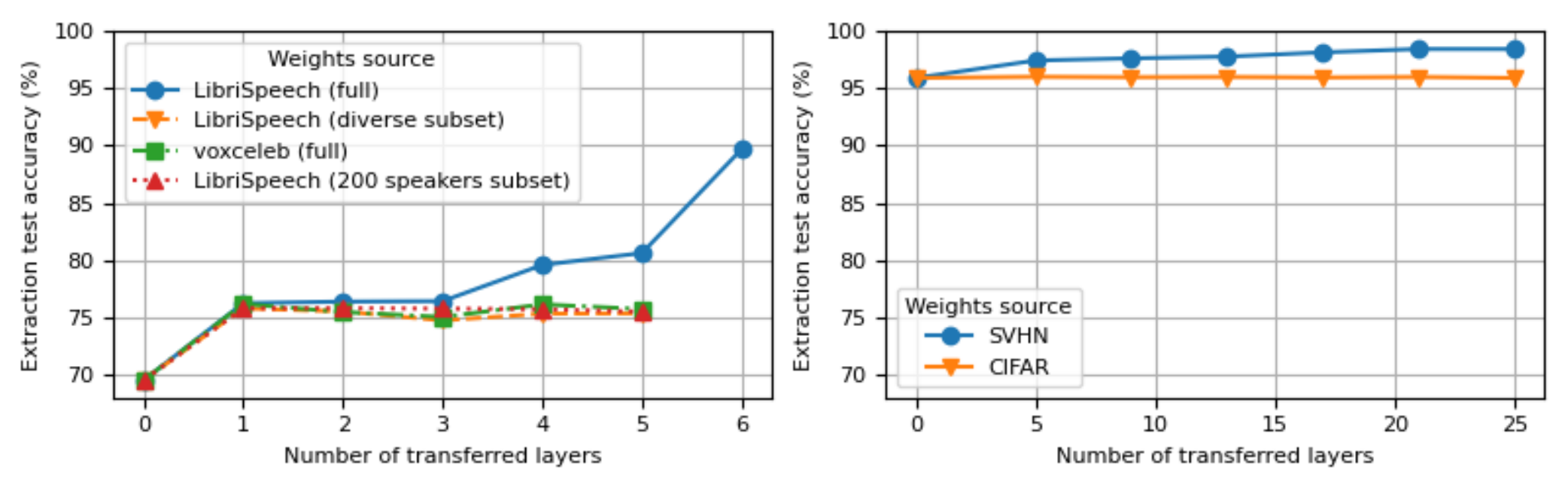}
\caption{Extraction test accuracy with respect to the number of transferred layers when extracting the \texttt{LibriSpeech KnaggCNN} victim with \texttt{VoxCeleb} (left) and the \texttt{SVHN WRN-28-10} victim with \texttt{CIFAR-100} (right): noticeable features include three test accuracy leaps on the \textit{blue} curve and only one on the others (left) and a steadily improving test accuracy as more layers are initialized no matter the source of the weights (right).}
\label{fig:layerwise}
\end{figure}

\begin{enumerate}
\item[{\bf E3.}] In the next experiment, we perform partial model extraction, {\em of a subset of the victim's layers}, rather than the entire architecture. We simulate this by giving the adversary access to the remaining layers' weight values. Practically, this means the adversary initializes its local copy with the subset of the weights it is provided (\ie the weights of the victim), and utilizes extraction to learn the rest. \textcolor{black}{We choose different subsets of the \texttt{LibriSpeech} dataset to train the victim (from which we choose different subsets to provide as {\em known weights} to the adversary); all models trained share the same architecture (refer to Appendix~\ref{A}).} They are trained on subsets of the victim's training dataset (\texttt{LibriSpeech}), and on the proxy dataset (\texttt{VoxCeleb}). By varying properties of the source task, we can see how the knowledge learned transfers in the extraction setting. We vary the number of layers made available to the adversary, starting with first layer only, then first two layers, and so until we reach the pre-softmax layer of the model. The attack otherwise runs as previously described, with the adversary querying the model with proxy data from the full \texttt{VoxCeleb} dataset in all cases. 

\item[\bf Results (E3):] Each curve in Figure~\ref{fig:layerwise} illustrates a sweep through the victim layers for a specific choice of victim weights (\ie those used to initialize the weights of the adversary, for extraction). We notice all curves present a characteristic test accuracy leap between layer zero (no access to victim layer weights at all) and layer one (weights of the sole first victim layer). Almost surprisingly this means that we are able to successfully extract and to recover the semantics of the first layer whatever task we use for extraction. Beyond the first layers, the blue curve that represents the extraction with \texttt{LibriSpeech} itself as source data keeps improving while others plateau. This demonstrates that no matter the task used to learn the weights for initialization, it does not help recover the subsequent layers that are thus more \textit{task-dependent}.
\end{enumerate}

For the purpose of comparison, we carried out the same experiment in the vision domain. We extract a \texttt{WideResNet}~\cite{zagoruyko2016wide} victim trained on \texttt{SVHN} following the procedure described in Appendix~\ref{A} with \texttt{CIFAR-100} as proxy dataset, and initialization weights from the same model trained on \texttt{SVHN} or \texttt{CIFAR-100}. Contrary to what happens in audio, results presented in Figure~\ref{fig:layerwise} (right) show how (a) extraction succeeds and how the accuracy improves quasi-linearly, and (b) how extraction is not so dependent on the task.

\begin{figure}[ht]
\centering
\includegraphics[width=\linewidth]{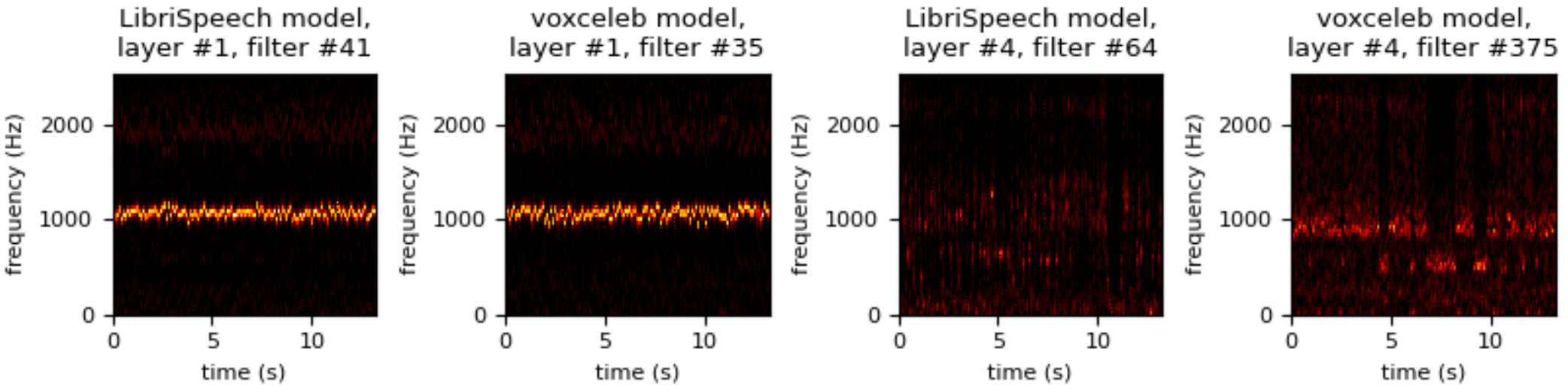}
\caption{Activation of chosen filters of the first and fourth layers of a \texttt{KnaggCNN} trained on \texttt{LibriSpeech} and \texttt{VoxCeleb}.}
\label{fig:visualization}
\vspace{-1em}
\end{figure}

\subsection{Exploring Task Dependence of Audio Features}
\label{sec:4.3}

A promise of DNNs is to leverage local features---learned by the model's early layers---to build more complex and non-local features in the later layers. In the case of vision, early filters of a convolutional NNs detect edges, corners, and textures in adjacent pixels that can be used as building blocks by subsequent layers. Similarly, early convolution layers of an audio model capture increasingly complex frequency band structures. However, the very notion of \textit{locality} is different in audio. For example, a pair of frequencies that differ by a factor of 2 are often both related and perceived as being similar\footnote{They can result as harmonic vibrations from the same source event.}.

We can visualize this by adapting gradient-based techniques such as the one of Simonyan \etal~\cite{simonyan2013deep} originally used in the vision domain to our audio models. Figure~\ref{fig:visualization} presents filters of the first and fourth layers of two \texttt{KnaggCNNs} trained on \texttt{LibriSpeech} and \texttt{VoxCeleb}; the choice of fourth layer is arbitrary. In both cases, the first layer seems to match very simple patterns such as a single frequency band which is reminiscent of Fourier Transform-based filters\footnote{Fourier Transform is a well known jack-of-all-trades representation in audio that is useful to a wide category of tasks.} whereas filters of the fourth layer appear to match more complex and broader patterns.

\begin{figure}[H]
\begin{center}
\includegraphics[width=0.5\textwidth]{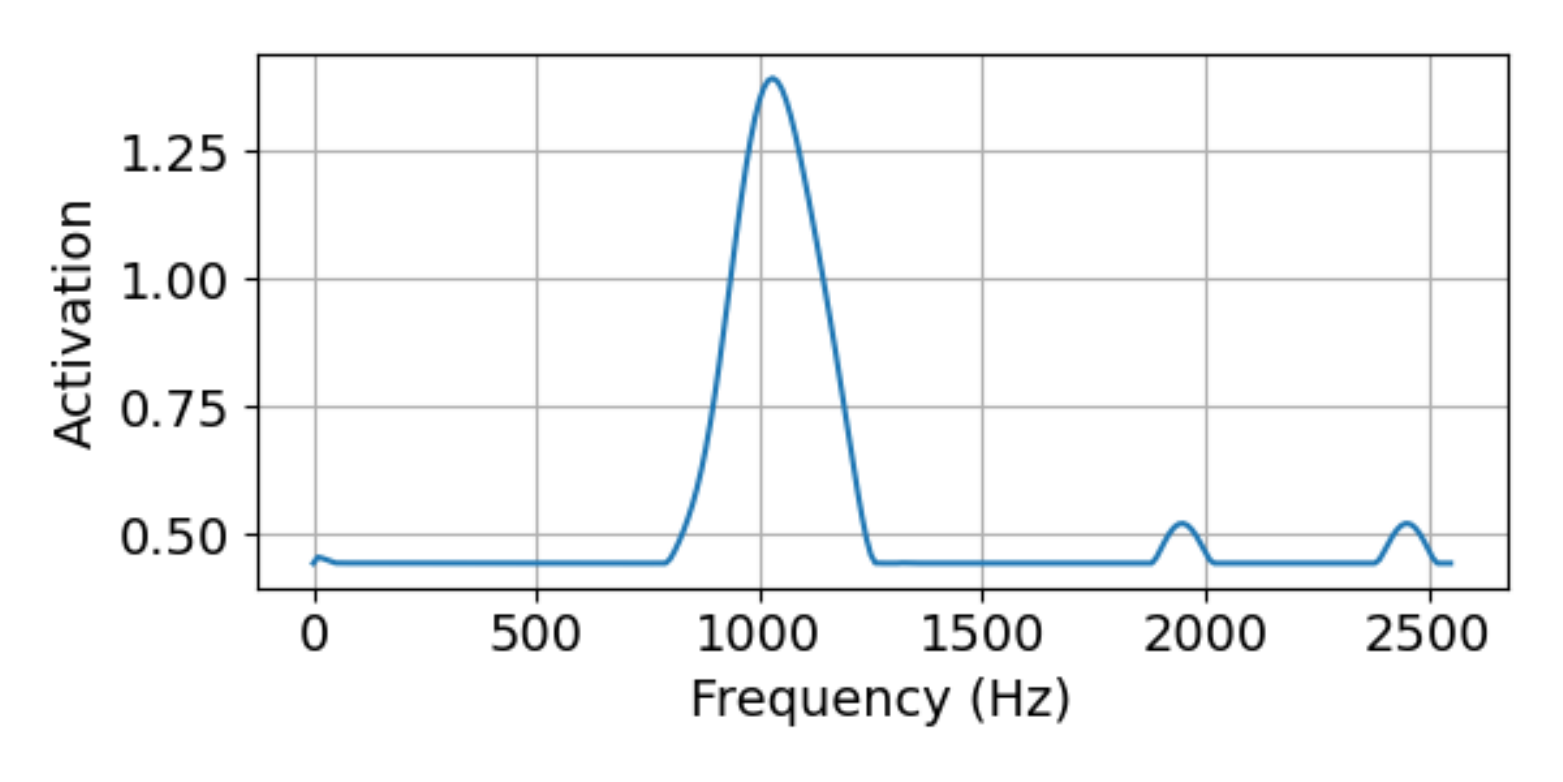}
\end{center}
\caption{Activation of a chosen filter (layer \#1, filter \#41) with respect to the frequency of the sine wave given as input to the model. Note the small spikes at harmonically-nearby frequencies.}
\label{fig:activation}
\end{figure}

This explains why the first layer of the \texttt{KnaggCNN} (in the layer-wise extraction experiment) is easy to extract with some prior audio knowledge. However, the problem of speaker identification which we are studying, requires deeper layers to reconstruct a representation of each speaker's frequency contour, a rather specific requirement. These task-specific features of the deeper layers make model extraction more difficult in audio without some prior knowledge.

\paragraph{Overcoming task-dependence with coverage.} Our layer-wise {extraction} experiment ({\bf E3}) demonstrates that there is a correspondence between task-specific features and the difficulty of extraction. Naturally, we posit that acoustically diverse queries would give more information, which would in turn increase the extraction test accuracy. Our experiments on coverage tend to second this intuition. We thus hypothesize for the rest of the paper that increased coverage of the adversary's queries enables them to better capture these task-specific features. This has proved fruitful to design our extraction techniques (evaluated in \S~\ref{sec:eval}).


\section{Generative Models for Extraction}
\label{sec:extraction:gan-synthesis}

Our results from \S~\ref{sec:motivation} motivate the shortcomings of porting prior work to the audio domain, and the task-dependence of learning from audio features. In this section, we wish to combine these insights into proposing a new extraction methodology: using generative models (trained on a proxy dataset) to synthesize samples. However, some important questions remain. 

\noindent{\bf Q1. What generative model do we use?} Generative models introduce trade-offs in various aspects including (a) fidelity of the generated audio, (b) computational resources required for synthesis, and (c) mode-covering vs. mode-seeking tendencies (and more generally, coverage of the true data distribution). Furthermore, the extraction query budget may be constrained either by the query interface provided by the victim or by the generative model's inference latency. 

\texttt{WaveNet}~\cite{oord2016wavenet} (an autoregressive model) can produce audio with high fidelity and therefore may require relatively few query responses from the victim. However, the latency of sampling from this model was prohibitive. We also investigated the effectiveness of several established audio synthesis approaches~\cite{jia2018transfer, hsu2018hierarchical, binkowski2019high}, but none of them proved suitable for our work due to various subtleties in their modeling approach that does not inherently lend itself to our proposed experiments, \eg in one case the model requires training a separate speaker encoder using an independent large database, another requires an unavailable conditioning signal, etc. Therefore, we use \texttt{WaveGAN} for our experiments~\cite{donahue2019adversarial}.

\noindent{\bf Q2. Why use synthetic queries?} Sampling synthetic queries has several advantages: (a) training generative models is an unsupervised task, and therefore the adversary does not need to interact with the victim to bootstrap the process, (b) once a generative model is trained, the adversary can potentially use it to obtain a more diverse set of queries than it originally had (using just the proxy dataset) by modifying the posteriors of the latent variables accordingly.

\noindent{\bf Q3. What sampling strategies can be used?} We propose and experiment with two techniques: (a) thresholding-based sampling, and (b) iterative sampling.

\begin{enumerate}
\item[(a)]{\bf Thresholding-based sampling:} Under the assumption where the adversary knows the true distribution of labels of the dataset used to train the victim ({\bf A3} from \S~\ref{sec:threat}), let $L=\{c_1, \dots, c_N\}$ denote the count/number of samples for each label $i \in \{1,\dots, N\}$ in the victim's training data. With this we can formalize two (query) thresholding strategies that limit query complexity:
\begin{enumerate}
\item[(i)]{\em Dynamic Thresholding:}  The adversary synthesizes queries such that the number of data points for label $i $ is $c_i' \leq \beta \cdot c_i$, where $\beta \geq 1$. Here, $\beta$ is known as the scaling parameter. 
\item[(ii)]{\em Static Thresholding:} Irrespective of the values of $c_i$, the adversary synthesizes queries such that the number of samples for each label $i$ is $c_i' \leq \alpha$. Here, $\alpha$ is the static threshold.
\end{enumerate}
The former is designed to increase the number of task-specific features, while the latter is designed to minimize computational overhead. In both settings, the adversary simply discards the rest of the queries synthesized once the threshold is reached. Our results in \S~\ref{sec:eval} are often comparable in both settings, meaning that even static thresholds alone can improve extraction accuracy. This suggests that {\bf A3} may not be needed in practice.
\item[(b)]{\bf Iterative sampling:} In the previous approach, the adversary generates and labels a potentially large number of (unbounded) samples such that it meets the required label thresholds. In this approach, the adversary generates a {\em fixed} set of samples (denoted $size$) per iteration (denoted $n$) and proceeds with static or dynamic thresholding. If at the end of query generation, the thresholds are not met, then the adversary proceeds with what is available. The approach is highlighted in Algorithm \ref{D-repeated-iterations} in the Appendix \ref{D}. We evaluated this approach only for $n=1,10$, and $size=300,000$. This implies that the maximum number of queries  generated is $N=size \times n$. Our aim is to reduce skewness in our extracted dataset. We report these results in \S~\ref{sec:eval}. In Table \ref{tab:thresholding_result}, we can see that the test accuracy for $n=1$ is comparable to the setting where we extracted the victim model using the $300,000$ samples in their entirety
\end{enumerate}

\section{Evaluation}
\label{sec:eval}

We wish to answer the following question: \textit{Does increasing the volume and coverage of data generated improve the ability to perform extraction?}

\begin{table}[ht]
\begin{center}
\begin{tabular}{lcccc}
\toprule
\bf Dataset & \bf Volume & \bf Coverage (\%) & \bf Extraction Test Accuracy (\%) \\
\midrule
\midrule
({\bf EB1}) \texttt{VoxCeleb}  & 92,607  &85.5  &  73.45  \\
({\bf EB2}) \texttt{LibriSpeech} &83,170    &100.0  & 94.31  \\
\midrule
({\bf GM1}) \texttt{VC-GAN}  &   300,000 &63.7  &  76.89 \\
({\bf GM2}) \texttt{LS-GAN}  &  300,000 &79.7  &  81.76  \\
\bottomrule
\end{tabular}
\caption{Extraction of a \texttt{LibriSpeech} classifier (victim) which achieves a test accuracy of 94.33\% using \texttt{LibriSpeech}, \texttt{VoxCeleb}, \texttt{VC-GAN} and \texttt{LS-GAN}. \texttt{LS-GAN} and \texttt{VC-GAN} are WaveGAN trained models and synthesized sample size is 300,000. Volume denotes number of extraction queries; Coverage denotes ratio between the number of unique labels predicted by the victim (using the proxy dataset) and those in the original victim dataset; Extracted Test Accuracy is accuracy of the extracted model. ({\bf EB1}) and ({\bf EB2}) trained for 80 epochs, ({\bf GM1}) and ({\bf GM2}) for 100 epochs.}
\label{tab:naive-extraction}
\end{center}
\end{table}

\noindent{\bf Setup:} As stated earlier, we train our \texttt{KnaggCNN} victim using \texttt{LibriSpeech} (refer Appendix~\ref{model:knagg}). The victim achieves a test accuracy of 94.33\%. Our learning-based adversary interacts with this victim in a black-box manner. We trained two generative models based on the \texttt{WaveGAN} architecture~\cite{donahue2019adversarial}, for reasons mentioned in \S~\ref{sec:extraction:gan-synthesis} (see Appendix~\ref{model:wavegan} for details). One generative model was trained with the proxy dataset \texttt{VoxCeleb}; we refer to this generative model as \texttt{VC-GAN}. Another genetative model was trained using the original \texttt{LibriSpeech} dataset; we refer to this generative model as \texttt{LS-GAN}. \texttt{VC-GAN} is used to evaluate the efficacy of our proposal, while \texttt{LS-GAN} is used evaluate the effectiveness of synthesized queries in approximating the true distribution of the data.

\noindent{\bf Baselines (Original Datasets):} ({\bf EB1}) When we extract the (\texttt{LibriSpeech} trained) \texttt{KnaggCNN} with the whole \texttt{VoxCeleb} dataset, we achieve a test accuracy of 73.45\% (see Table \ref{tab:naive-extraction}). ({\bf EB2}) To see how much room for improvement there is, we also report a second baseline--an adversary extracting \texttt{KnaggCNN} with the \texttt{LibriSpeech} dataset. Here, we achieve an accuracy of 94.31\% which is almost the same as the victim's test accuracy (which is expected). 

\begin{table}[ht]
\begin{center}
\begin{tabular}{clcccc}
\toprule
\bf Iteration (n) & \bf Thresholding Parameter & \bf Volume & \bf Coverage (\%) & \bf Extraction Test Accuracy (\%) \\
\midrule
\midrule
1 & $\beta=1$  &\raggedleft 37,563 &64.2 &39.84 \\
& $\beta=10$  &\raggedleft 143,395  &64.5 & 76.27 \\
& $\alpha=500$ &\raggedleft 96,076 &64.2 &65.31  \\ 
& $\alpha=1000$ &\raggedleft 137,449 &64.5 & 72.79 \\
\midrule
10 & $\beta=1$  &\raggedleft 60,607  &78.3 &57.82 \\
& $\beta=10$  &\raggedleft 231,270  &77.9 & 84.12  \\
& $\alpha=500$ &\raggedleft 164,077  & 78.3 &80.42 \\
& $\alpha=1000$ &\raggedleft 222,593 &77.9 & 84.90 \\
\bottomrule
\end{tabular}
\caption{Extraction of a \texttt{KnaggCNN} using \texttt{VC-GAN} with $n=1$ (Top) and $n=10$ (Bottom), with varying threshold settings. Volume denotes the number of extraction queries, Coverage denotes the ratio between the set of unique labels predicted by the victim (using the proxy dataset) and the unique labels in the original victim dataset. All numbers obtained after 80 epochs of training.}
\label{tab:thresholding_result}
\end{center}
\end{table}

\noindent{\bf Baselines (Generative Models):} We set an upper-bound of 300,000 synthetic queries for this experiment. 

({\bf GM1}) Observe that extracting the victim with \texttt{VC-GAN} results in a model with test accuracy of 76.89\%. This clearly outperforms extraction with \texttt{VoxCeleb} (73.45\%) suggesting that increasing the volume of the dataset has a quasi-linear relationship with the extraction accuracy. Although extraction with \texttt{VoxCeleb} has higher coverage than using \texttt{VC-GAN}, \texttt{VC-GAN} can often compensate for degree of coverage with higher volume. A natural next step would be to increase volume further, but we found that this ultimately degrades extraction accuracy due to high skewness in the data. 

({\bf GM2}) Extracting the victim with \texttt{LS-GAN} results in a model with test accuracy of 81.76\% (much lower than that obtained by using the original dataset), this can be attributed to poor coverage of victim's speakers' distribution, by iteratively sampling the \texttt{LS-GAN} to maximize speaker coverage, the extraction results is often comparable to using the original dataset. 

To understand the interplay between volume and coverage and its influence on extraction, we utilize the iterative sampling strategy as outlined in \S~\ref{sec:extraction:gan-synthesis}. As stated earlier, we set $n=1,10$ and $size=300,000$ in these experiments. In Table~\ref{tab:thresholding_result}, we measure both volume and coverage. Increasing both volume and coverage results in the best setting for extraction of 84.90\% (refer to the row with $\alpha=1000$, $n=10$) but comes at a higher cost (\ie $N=3,000,000$). By judiciously synthesizing queries from the \texttt{VC-GAN} model, we increase the speaker coverage and thus the extraction accuracy. 

\begin{figure}[ht]
     \begin{subfigure}
    \centering
    \includegraphics[width=0.5\textwidth]{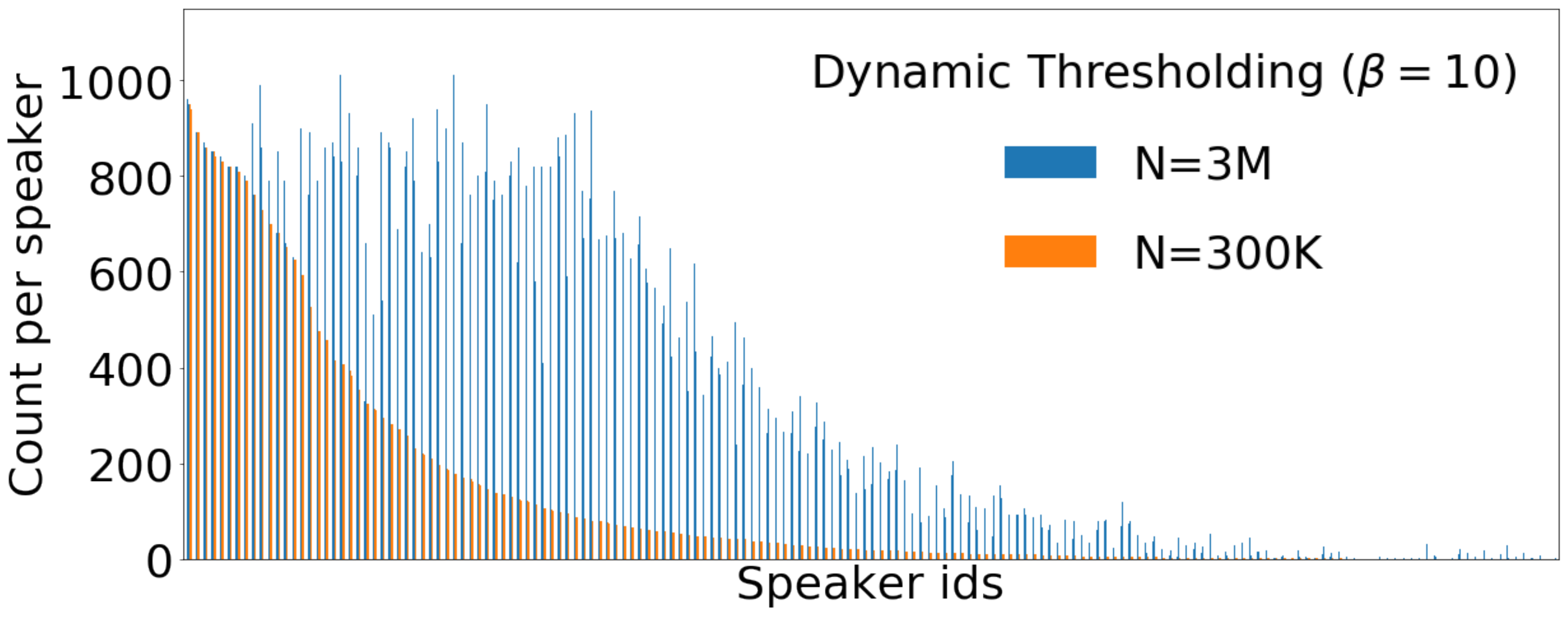}
    \end{subfigure}
    \hfill
    \begin{subfigure}
    \centering
    \includegraphics[width=0.5\textwidth]{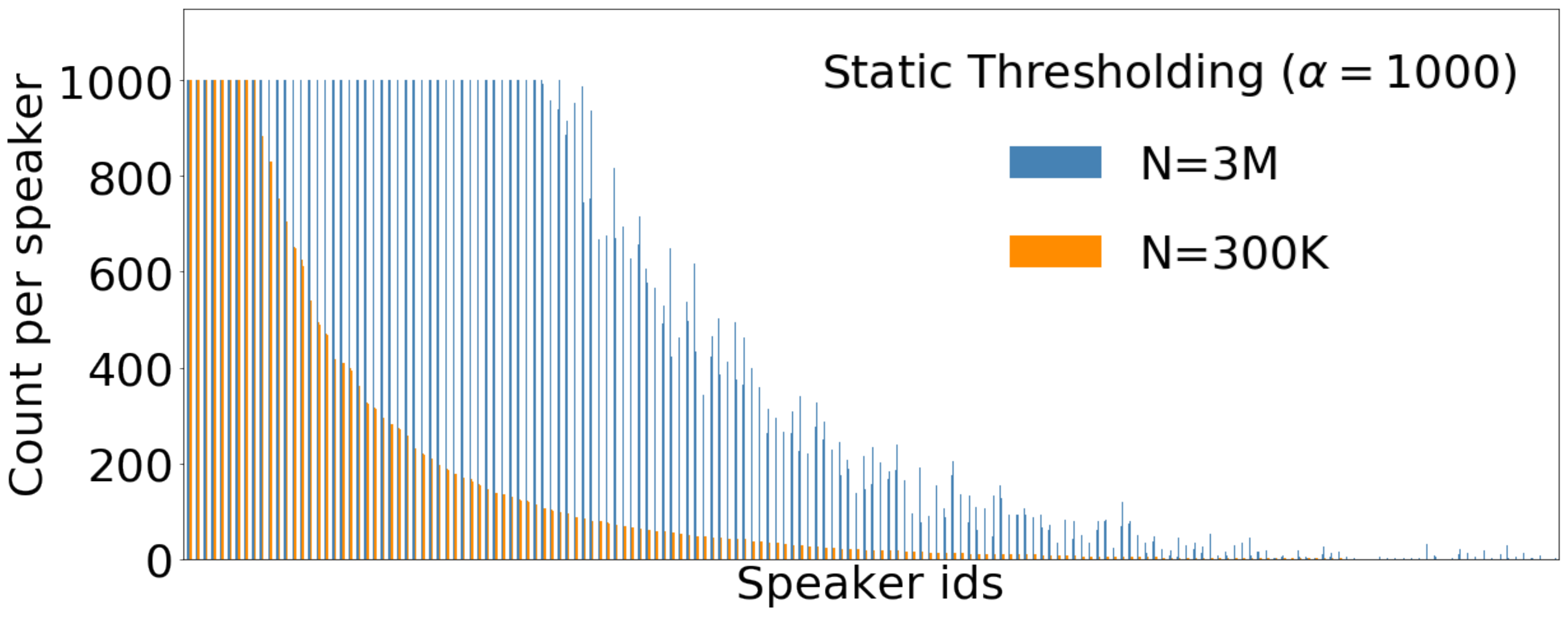}
    \end{subfigure}
   \caption{The number of samples per speaker increases with the number of queries sampled from \texttt{VC-GAN}, with static and dynamic thresholding, thus improving extraction accuracy. (Left) shows Dynamic thresholding with $\beta=10$ and (Right) shows the static thresholding at $\alpha=1000$.}
    \label{fig:vox_static_dyn}
\end{figure}

From Figure~\ref{fig:vox_static_dyn}, one can see that as the value of $n$ increases, coverage also increases in both the static and dynamic variants of iterative thresholding. This consequently leads to better extraction, as reported in Table~\ref{tab:thresholding_result}. While these trends suggest that iterative sampling is useful for extraction, we were unable to increase $N$ beyond $3,000,000$ due to computational reasons (generating samples from the generative model is computationally expensive). An adversary also faces diminishing returns as the query complexity of the attack increases. As expected, iterative sampling increases volume and coverage, resulting in better extraction.

\section{Limitations}
\label{sec:limitations}

Our experiments involve extracting a model trained with \texttt{LibriSpeech} with variants of \texttt{VoxCeleb}; the reverse setting is detailed in Appendix~\ref{E}. The underlying assumption is the existence of a {\em suitable} proxy dataset that can bootstrap extraction. However, properties (i.e., number of samples, desired coverage, nature of acoustic diversity etc.) of said proxy dataset are not defined, and are often tied to (a) the task at hand, and (b) the original dataset used for extraction. Furthermore, training generative models incurs extensive computational and environmental costs. Additionally, query complexity of our extraction approach can further be reduced by using generative models that are more expressive, and can be conditioned on various acoustic features. Addressing these limitations is future work.

\section{Conclusions}

Model extraction in audio has proven challenging. Certain features learned by audio classifiers appear to be more task-dependent and increase the query complexity of model extraction. We thus leveraged generative models to increase both the volume and coverage of data the adversary queries the model with. 
To systematize our understanding of this area, we find that comparing model extraction across a wide range of audio settings would be an interesting problem to tackle in future work (e.g. with more specialized architectures~\cite{ravanelli2019speaker}, and multi-task datasets~\cite{lee2019label} that might lead to more general features at more layers), as well as integrating any advances in generative modeling of audio.

\newpage
\bibliographystyle{unsrt}
\bibliography{main}

\newpage
\appendix
\section*{Appendix}
\section{Datasets \& Models}
\label{A}

We describe various technical details related to the models and datasets we consider in our experiments.

\subsection{Audio Datasets}

Salient features of all datasets used are consolidated in Table~\ref{tab:A-dataset-description}

\begin{table}[H]
\begin{center}
\begin{tabular}{ccc}
\toprule
{\bf Dataset}  & \bf {\bf \# Samples} & \bf {\# Speakers}\\ 
\midrule
\midrule
\texttt{LibriSpeech}  & 137,779 &1251\\
\texttt{VoxCeleb} &153,516   &1251 \\
\bottomrule
\end{tabular}
\caption{Dataset Description}
\label{tab:A-dataset-description}
\end{center}
\end{table}

\subsubsection{LibriSpeech} 
\label{data:librispeech}

\texttt{LibriSpeech}~\cite{panayotov2015librispeech} is an ASR corpus based on public domain audiobooks. It consists of 1000 hours of read 16KHz English from the \texttt{LibriVox} corpus of audiobooks. 

\begin{figure}[H]
\centering
\includegraphics[width=0.8\linewidth]{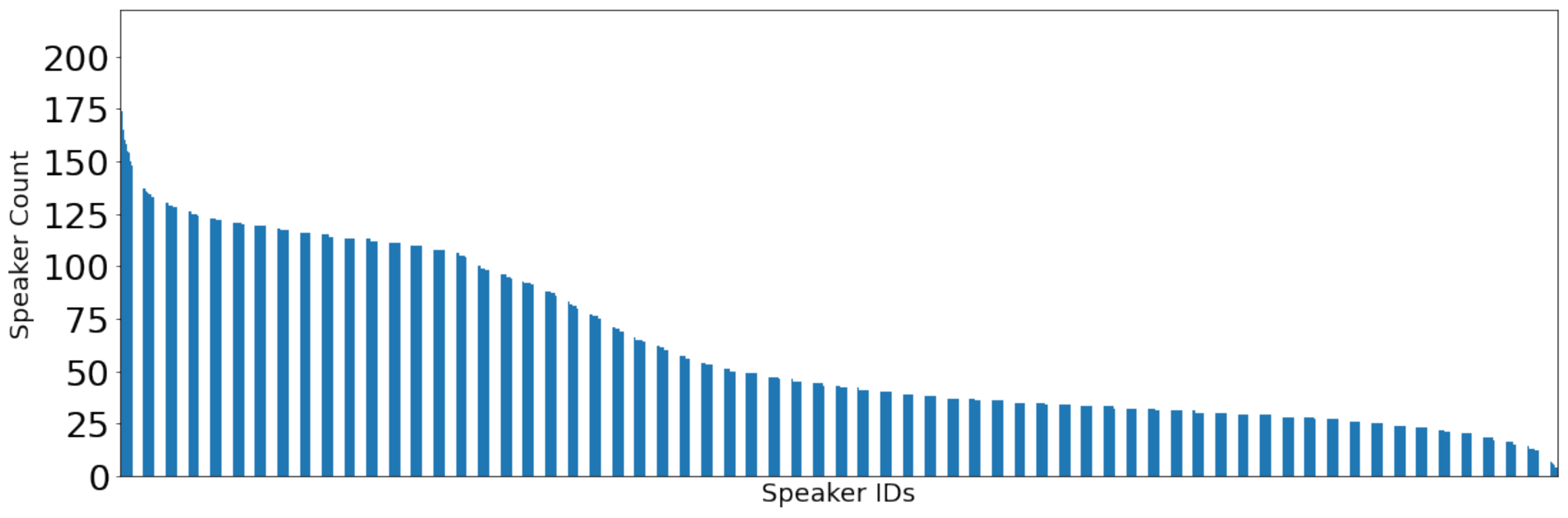}
\caption{Distribution of number of samples per speaker for \texttt{LibriSpeech}. The x-axis represents the speaker IDs (1251 speakers) and y-axis is the number of samples per speaker. 
}
\label{fig:A-speaker_distribution}
\end{figure}

In our experiments, we used the subsets \texttt{LibriSpeech-clean}, \texttt{train-clean-360} (961 speakers), \texttt{train-clean-100} (251 speakers), \texttt{test-clean} (40 speakers), and \texttt{dev-clean} (40 speakers). We combined these subsets and omitted speaker $60$ in order to maintain a total size of $1251$, which is the size of our proxy dataset \texttt{VoxCeleb}. We custom split the dataset and used 83,170, 27,562 and 27,047 samples for training, testing and validation data and maintains the proportion of speakers in the three parts. Figure~\ref{fig:A-speaker_distribution} shows the \texttt{LibriSpeech} speaker distribution with the 4 subsets combined.

\subsubsection{VoxCeleb} 
\label{data:voxceleb}

\texttt{VoxCeleb}~\cite{nagrani2020voxceleb} is an audiovisual dataset consisting of short clips of human speech extracted from interview videos uploaded to YouTube. The full version of the dataset is very large, with more than 7,000 speakers, one million utterances, and 2,000 hours of speech. It encompasses many different situations and environments because it was recorded in the wild, as opposed to datasets recorded in a controlled manner in a studio. We use a subset of \texttt{VoxCeleb} that contains the audio speech of 1251 different speakers. As with LibriSpeech, the data is organized into FLAC audio files at 16kHz 16bits.
We used 96607, 30705, and 30204 samples for training, testing, and validation data and maintains the proportion of speakers in the three parts.

In essence, \texttt{VoxCeleb} and \texttt{LibriSpeech} are similarly structured; both roughly have the same number of speakers and a comparable number of samples, except \texttt{VoxCeleb} has more diverse data with speakers spanning a wide range of different ethnicities, accents, professions, and ages.

\subsection{Audio Models}
\label{models}

\subsubsection{Knagg CNN} 
\label{model:knagg}

This model, hereafter referred to as \texttt{KnaggCNN}, was adopted from an online post~\cite{knagg_building_2018}, and was chosen as a relatively simple yet effective network for classification based on raw audio i.e. digital audio signals represented as arrays of numbers. It is a CNN with 4 1D convolutions, batch normalization and (optional) spatial dropout, followed by two fully connected layers with an embedding size of 64 (we used 128 in our experiments). The number of filters per network layer ranges from 128 to 512, and we used a filter size of ($32 \times 1$) for the first layer, which is both small enough to allow sample-level features and large enough to allow some frame-level features. We train this using the Adam optimizer, with learning rate $\eta =0.001$, and standard \texttt{pytorch} parameters $\beta_0$ and $\beta_1$, with no dropout. We augment data by randomly selecting a window of the original audio with the correct input dimension (about 1 seconds). The original code of this model is available on github at \url{https://github.com/oscarknagg/voicemap}. The version we use is available in our repository (refer Appendix~\ref{F}).

\subsubsection{WaveGAN} 
\label{model:wavegan}

Our experiments in  \S~\ref{sec:eval} utilizes \texttt{WaveGAN}~\cite{donahue2019adversarial} to generate synthetic samples. \texttt{WaveGAN} is an unsupervised generative adversarial network (GAN) used to synthesize raw waveform audio. The \texttt{WaveGAN} architecture is based on DCGAN, which popularized the use of GANs for image synthesis. We trained our \texttt{WaveGAN} models with the default hyperparameters (see Table~\ref{tab:A-wavegan_hyperparam} for some of parameter settings) on \texttt{VoxCeleb} and \texttt{LibriSpeech} datasets (all 1251 speakers). Both models were trained for approximately 170,000 steps and 1 second audio samples were synthesized.

\begin{table}[H]
\begin{center}
\begin{tabular}{ll}
\toprule
\bf Hyperparameters & \bf Value\\
\midrule
\midrule
Normalize the training examples &False \\
Number of audio channels to generate & 1 \\
Overlap ratio [0, 1) between slices & 0.0 \\ 
Use zero-padded partial slices from the end of each audio file & False \\ 
Only use the first slice each audio example & True \\ 
Number of audio samples per second &16000 \\
Number of audio samples per slice & 16384 \\
Batch size & 64 \\
Number of dimensions of the latent space & 10 \\
Dimensionality multiplier for model of G and D &64 \\
Enable batchnorm &False \\
Length of 1D filter kernels & 25 \\
Discriminator number of updates & 5 \\
Radius of phase shuffle operation & 2 \\
Use post-processing filter & False \\
Length of post-processing filter for DCGAN &512 \\
Generator upsample strategy & zeros \\
Which GAN loss to use & wgan-gp \\
\bottomrule
\end{tabular}
\end{center}
\caption{WaveGAN Hyperparameter Settings}
\label{tab:A-wavegan_hyperparam}
\end{table}
 
\subsection{Vision Models}

\subsubsection{WideResNet} 
\label{model:wrn}
 
WideResNets (WRNs)~\cite{zagoruyko2016wide} are an improvement of ResNets~\cite{he2016identity} that use wider and fewer layers that the original. Like ResNets they rely on identity shortcuts that among other benefits help alleviate the well known issue of vanishing gradients. WRNs come in various sizes, we use WRN-28-10 in this paper. We train it with bare SGD, learning rate $\eta = 0.1$, mini-batch size 128, momentum 0.9, weight decay 0.0005 and we train for 200 epochs. In addition, we reduce the learning rate by 80~\% after 60 and 120 epochs as suggested in \cite{zagoruyko2016wide}. We also use dropout with a probability 0.3 in the wide residual blocks. We do not use any data augmentation.

\section{Differences between Audio and Images}
\label{B}

We highlight salient differences between learning from images and audio relevant to model extraction.

{\bf D1. Input Representations:} One natural representation for audio signals is the digital (``raw'') waveform itself that can be encoded as a 1D tensor, where each element in the tensor represents a quantized, sampled amplitude of the analog waveform. For example, one second of audio, sampled at 16KHz, would correspond to a 1D tensor of length 16,000. This is analogous to representing images with 2D tensors corresponding to the raw output of a camera. However, information is embedded in raw audio quite differently from how it is embedded in images. 
As a brief intuition for this, consider that the cochlea in the inner ear plays an important role in hearing by separating the various frequencies that compose the sounds we perceive. This is in contrast to the role that the frequency composition of light plays in the vision: we can see a lot of the world in terms of intensity (e.g. many gray-scale images are still relatively interpretable). The photo-receptors responsible for color vision are indeed activated by broad frequency bands in comparison to hearing. 
Reliance on frequency composition of audio semantics has led to 2D representations of audio, based on spectral representations (e.g. as obtained through Fourier transforms). STFT is the operation involved in computation of spectrograms that are 2D representations of audio accounting for evolution of the frequency distribution of the signal with respect to time. CQT is a variant of Fourier transform that in essence uses a log-space to represent frequencies and computes energies across progressively smaller temporal windows for increasingly higher frequencies. MFCC is the result of applying filter banks and Discrete Cosine Transform (DCT) to STFT to decorrelate the features. It has been heavily used in speech processing as a set of engineered features as it allows to separate the content---i.e., what a voice says---from the contour---i.e., the features of that voice. Magnitude spectrograms can be efficient, and allow the application of certain image-processing techniques, but also ignore phase information. While the perceptual impact of relative phase across time is not well understood, generating audio with randomized phase easily demonstrates that it is one of the characteristics that affect timbral qualities of the sound (e.g. the quality of a violin versus a flute, or of one speaker’s voice compared to that of another). The loss of this information may thus be detrimental to speaker identification applications for instance.
Despite their limitations, magnitude spectrograms are a very common representation of features for audio processing, and their 2D structure allows them to be processed somewhat like images. However, they are usually not processed \textit{exactly} like 2D images...

\vspace{1mm}
{\bf D2. Spectrograms are not images:} The two axes of a spectrogram do not each convey the same information. In the case of images, both axes are spatial dimensions that contain features at the same scale. Furthermore, for most visual data, axes could theoretically be reversed without any loss of information, even though it is not desirable in practice. On the contrary, spectrograms are composed of axes representing time and frequency, two entirely different dimensions. This implies, for example, that it does not make much sense to use typical square filters~\cite{dieleman2014recommending}. 

\vspace{1mm}
{\bf D3. Features are inherently different:} Local features can be used in both domains to reconstruct more abstract and non-local features. In the case of vision, the early stages of a CNN detect edges, corners, and textures across adjacent pixels that can be used by subsequent layers, e.g. to detect objects. Similarly, the first convolution layers of an audio model capture increasingly complex frequency band structures. However, the very notion of \textit{locality} is different in audio. For example, a pair of frequencies that differ by a factor of 2 are often both related (i.e., they can result as vibrations from the same source event), and they are perceived as being very similar as well. The result of this is seen in Figure~\ref{fig:activation}. On the classification problems we studied, deeper layers reconstruct a representation of the specificity of each speaker's frequency contour.

\vspace{1mm}
{\bf D4. Audio signals are {\em transparent}:} Overlapping sound objects do not mask one another, but are added together. That is, a single sample (one scalar value from the representation of the sound after sampling) can result from several sound objects at the same time. On the other hand, images are not transparent in the sense that any single pixel usually corresponds to a single visual object closest to the viewer along that direction; overlapping objects typically result in full or partial occlusion. 

\vspace{1mm}
{\bf D5. Time is a non-invertible dimension:}.
Time encodes a sense of causality which has two important consequences. First playing a sound from the end to the beginning alters its semantic meaning. This characteristic is however not audio specific: it is very common for \textit{non-natural} artifacts in vision such as writing. Flipping a character along a spatial dimension alters its meaning (e.g. a p reversed is a q). Second and more fundamentally, a given sample of a signal cannot depend on the following ones as it would break causality. This notion is absent in vision -- except maybe for OCR though this has more to do with textual data being a time series. 
 \section{Visualization \& Matching of Audio Filters}
 \label{C}
 
\subsection{Technical Details}

The optimized raw audio signals that correspond to the visualization of given filters in the architecture take the form of arrays of numbers between -1 and 1. This constraint is enforced throughout the algorithm by clamping after each improvement and is key to avoid any divergence.

We initialize the signals with uniform random arrays. At each step we compute the mean activation of the target filter by passing the signal being optimized as input to the model. This value serves as the objective of the optimization. The input is updated by simply backpropagating and taking an optimization (Adam) step in the direction of the normalized gradient with a step size equal to the unit.

To improve the resolution we use several rounds of optimization with increasing size of the optimized input, a technique known as "Taking it up an octave"~\cite{deepdream_tutorial}. We use 100 steps per (so called) octave, and 5 octaves total with a ratio 1.8 between two consecutive octaves (proper octaves would require a ratio of two, the ratio is treated as an hyperparameter here). We start at $1.8^{-2}\times$ the normal input size of the model and output an optimized audio that is $1.8^2\times$ larger than the normal input to the model.
 
\subsection{Matching Procedure}
 
When we train the same architecture with the same hyperparameters on two different datasets, there is no reason parallel filters in the two models will learn the same features. In the case of domain specific features, even if we expect to find the same or related filters in a given layer they will likely be permuted. Thus, to determine whether we have similar filters in a given layer---which is fundamental if we seek to analyse what this layer computes---we first need to match the filters by pairs before actually doing any pairwise comparison.

For each filter of one model, we select the filter in the other model that minimize the cosine distance in a specific feature space. We notice that there exist specific activations containing very low frequency components that are easily matched to any filter and thus use a custom filtering method to remove them from the comparison. In practice we use a 10 filters Mel-scale filter bank to compute the features of our 10-dimensional feature space.

\begin{figure*}[ht]
     \centering
     \includegraphics[width=\linewidth]{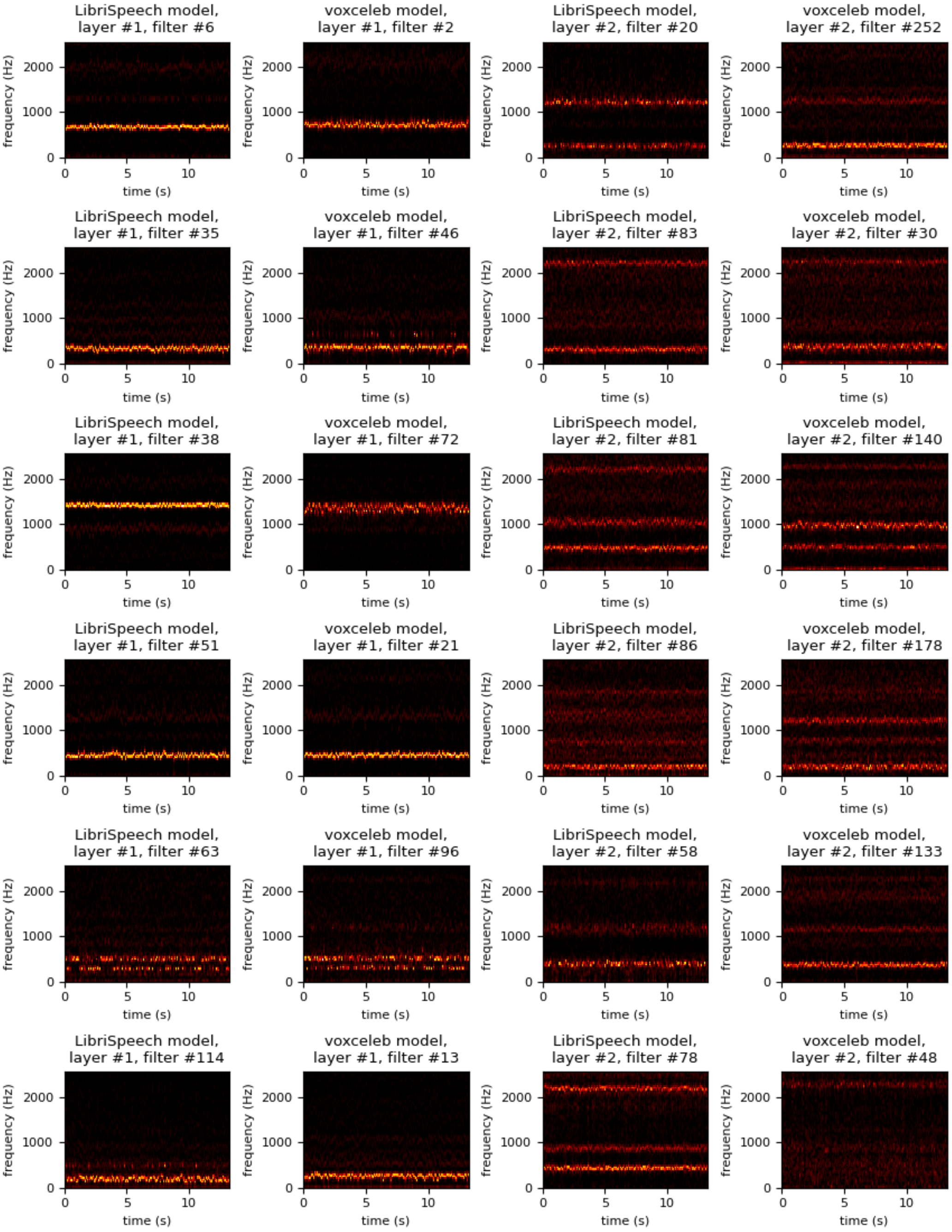}
     \caption{Activations of the first (first and second columns) and second (third and fourth columns) layers of a Knagg CNN trained on \texttt{LibriSpeech} (first and third columns) and \texttt{VoxCeleb} (second and fourth columns) matched (first column with second, third with fourth) by minimizing the cosine similarity of Mel-scale filter bank coefficients (10 filters).}
     \label{fig:C-more_activations}
 \end{figure*}

\begin{figure*}[ht]
     \centering
     \includegraphics[width=\linewidth]{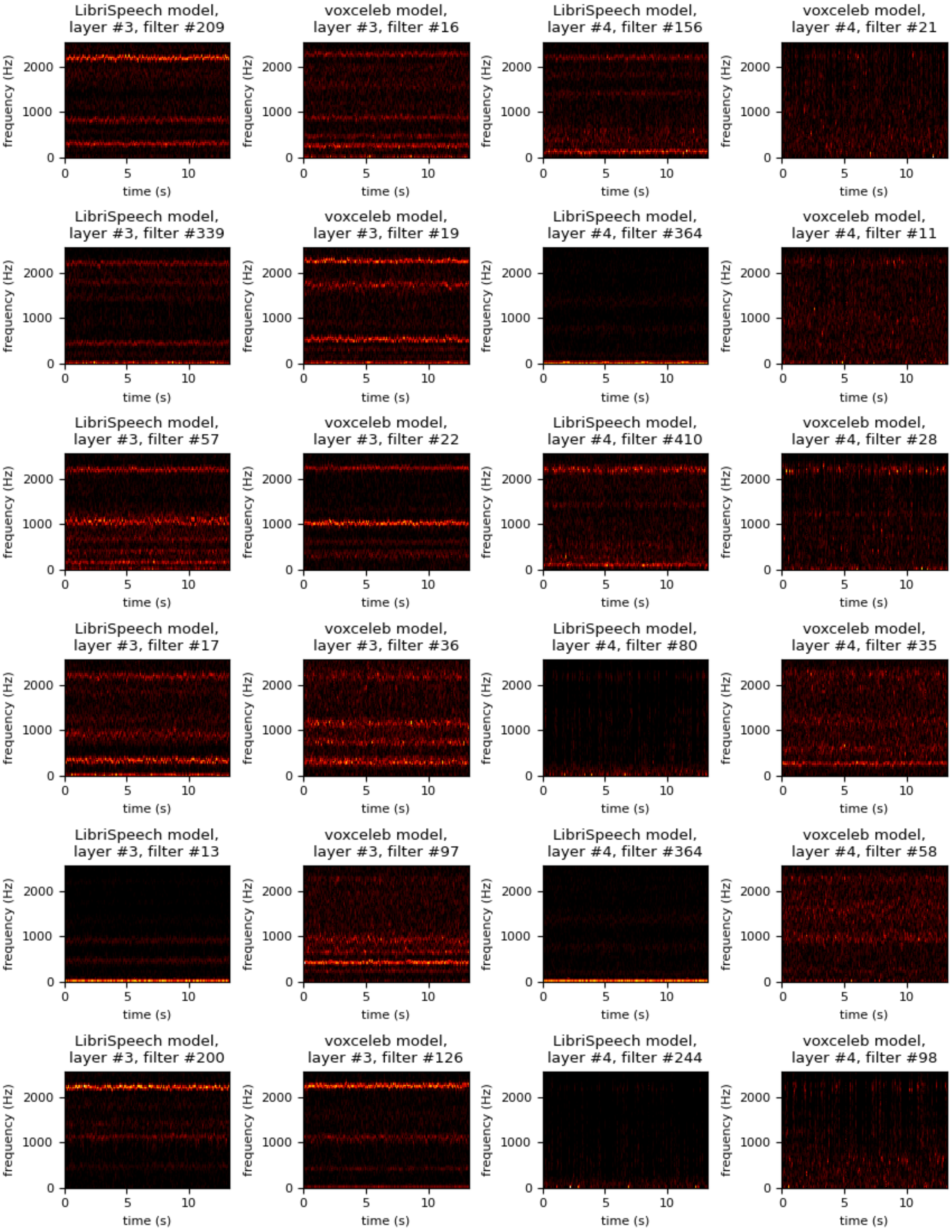}
     \caption{Activations of the third (first and second columns) and fourth (third and fourth columns) layers of a Knagg CNN trained on \texttt{LibriSpeech} (first and third columns) and \texttt{VoxCeleb} (second and fourth columns) matched (first column with second, third with fourth) by minimizing the cosine similarity of Mel-scale filter bank coefficients (10 filters). Note that matching layer \#4 does not work well as features are more task dependent.}
     \label{fig:C-more_activations2}
 \end{figure*}

\begin{figure*}[ht]
     \centering
     \includegraphics[width=0.5\linewidth]{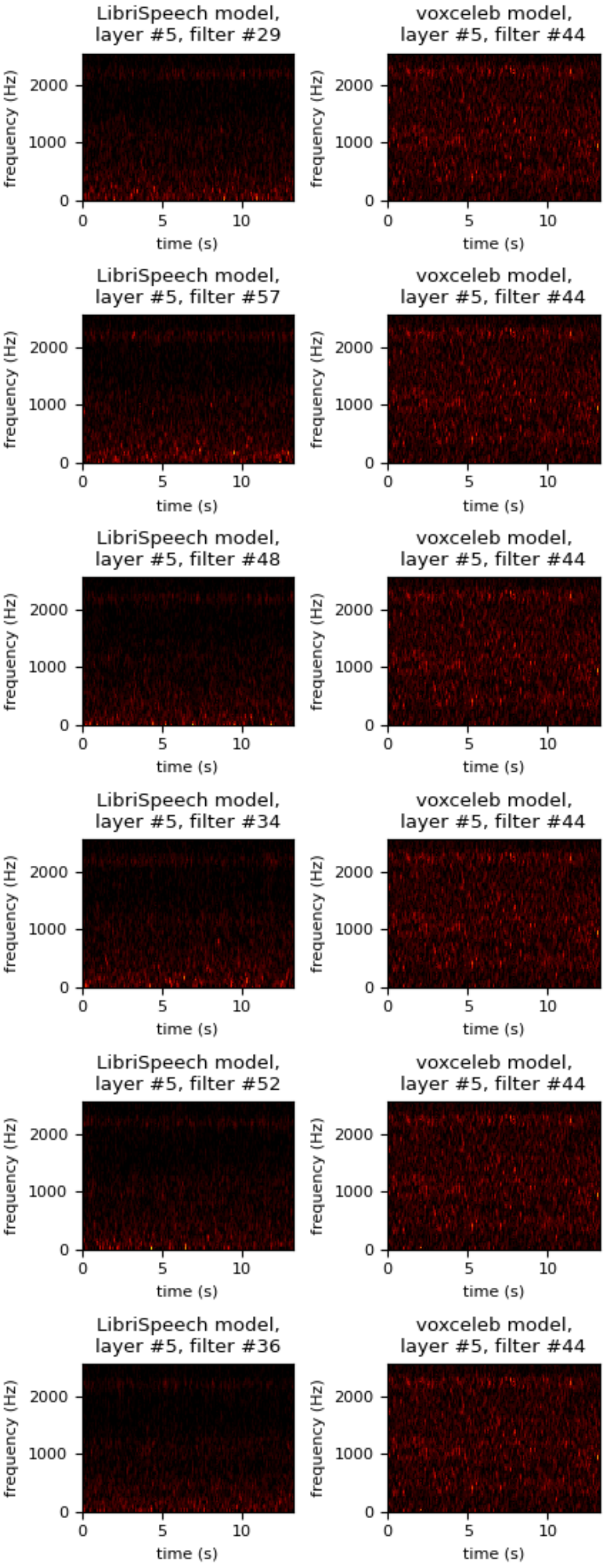}
     \caption{Activations of the fifth layers of a Knagg CNN trained on \texttt{LibriSpeech} (first column) and \texttt{VoxCeleb} (second column) matched by minimizing the cosine similarity of Mel-scale filter bank coefficients (10 filters). Note that matching layer \#5 does not work well as features are more task dependent.}
     \label{fig:C-more_activations3}
 \end{figure*}

A randomly selected subset of approximately matched output pairs is available in Figures~\ref{fig:C-more_activations}-\ref{fig:C-more_activations3}.
 
\section{Algorithm for Repeated Queries}
\label{D}

Refer Algorithm~\ref{D-repeated-iterations}. Line 5 synthesizes audio samples from the trained GAN models with $size$ denoting the number of samples to synthesize and line 6 labels these synthesized samples using the oracle model.

\begin{algorithm}[H]
\caption{Querying via repeated iterations}
\label{D-repeated-iterations}
\begin{algorithmic}[1]
\STATE count=1
\STATE dynamicDataset=0
\STATE staticDataset=0
\WHILE{count $\leq$ n}
    \STATE synthesizeAudio(size) 
    \STATE queryOracle() 
    \STATE dynamicDataset += dynamicThreshold()
    \STATE staticDataset += staticThreshold()
    \STATE count +=1 
\ENDWHILE\\
runExtraction(dynamicDataset)\\
runExtraction(staticDataset)
\end{algorithmic}
\end{algorithm}

\section{Extraction Experiments}
\subsection{VoxCeleb Oracle with \texttt{LS-GAN}}
\label{E}

Here, we reverse the setup, i.e., we extract a classifier trained on \texttt{VoxCeleb} using \texttt{LibriSpeech}. We report these results in Tables~\ref{tab:E-naive-extraction-reverse} and~\ref{tab:E-thresholding_result-reverse}. Unlike Table~\ref{tab:naive-extraction} in \S~\ref{sec:eval}, Table~\ref{tab:E-naive-extraction-reverse} shows that extracting the oracle \texttt{VoxCeleb} with \texttt{ LS-GAN} is not very effective and therefore the use of the proxy dataset, \texttt{LibriSpeech}, is best for all the extraction strategies studied, i.e. naive, thresholding, iterative sampling. We suspect that this is due to the poor performance of the \texttt{VoxCeleb} oracle classifier itself (64.89\%) as opposed to the \texttt{LibriSpeech} oracle (94.33\%), or it could simply be due to the nature of the dataset. The same conclusion can be drawn with table ~\ref{tab:E-thresholding_result-reverse}. In the future, we will evaluate the reverse setting with a more powerful oracle classifier.

\begin{table}[h]
\begin{center}
\begin{tabular}{l c c c c} \toprule
\bf Dataset & \bf Volume & \bf Coverage (\%) & \bf Extraction Test Accuracy (\%) \\
\midrule \midrule
VoxCeleb (baseline)  &92,607   &100.0  &  63.14 \\
LibriSpeech &83,170    &99.12  & 51.57 \\
VC-GAN  &   300,000 &88.4   &  47.99 \\
LS-GAN  &  300,000 &81.9 &  29.88 \\
\bottomrule
\end{tabular}
\caption{Extraction of a \texttt{VoxCeleb} classifier which achieves a test accuracy of 64.89\% using \texttt{VoxCeleb}, \texttt{LibriSpeech}, \texttt{VC-GAN} and \texttt{LS-GAN}. \texttt{LS-GAN} and \texttt{VC-GAN} are WaveGAN trained models and synthesized sample size is 300,000. Volume denotes the number of extraction queries, Coverage denotes the percentage of speakers from the proxy dataset labelled by the oracle, Extraction Test Accuracy is accuracy of the extracted model.}
\label{tab:E-naive-extraction-reverse}
\end{center}
\end{table}

\begin{table}[h]
\begin{center}
\begin{tabular}{l c c c c c} \toprule
\bf Iteration & \bf Thresholding Parameter & \bf Volume & \bf Coverage (\%) & \bf Extraction Test Accuracy (\%) \\
\midrule \midrule
1 & $\beta=1$  &\raggedleft 37,495  &81.7 &7.4 \\
& $\beta=10$  &\raggedleft 143,691 &81.7  &18.73  \\
& $\alpha=500$ &\raggedleft 163,742 &81.7 &14.44  \\ 
& $\alpha=1000$ &\raggedleft 137,940 &81.7 & 18.93 \\
\midrule
10 & $\beta=1$  &\raggedleft 60,411  &92.6 &11.72 \\
& $\beta=10$  &\raggedleft 230,838 & 93.2  & 34.50 \\
& $\alpha=500$ &\raggedleft 163,742 &92.6 &26.70 \\
& $\alpha=1000$ &\raggedleft 222,305 &93.2 &34.00 \\ 
\bottomrule
\end{tabular}
\caption{Extraction of a \texttt{VoxCeleb} classifier using \texttt{LS-GAN} with $n=1$ and $n=10$, with varying threshold settings. Volume denotes the number of extraction queries, Coverage denotes the percentage of speakers from the proxy dataset labelled by the oracle, Extraction Test Accuracy is accuracy of the extracted model.}
\label{tab:E-thresholding_result-reverse}
\end{center}
\end{table}

\section{Code \& Experimental Platform}
\label{F}

All extraction experiments from \S~\ref{sec:eval} were carried out with 4 Azure Standard NC24r servers equipped with one-half K80 card, 48GiB memory and 24 vCPU. For experiments in \S~\ref{sec:motivation}, we utilized 2 workstations equipped with a Nvidia Titan Xp GPU, 12 Intel(R) Xeon(R) W-2133 CPU @ 3.60GHz and 15.4GiB memory each; 1 server equipped with 4 Nvidia GeForce RTX 2080 Ti, 20 Intel(R) Xeon(R) Silver 4210 CPU @ 2.20GHz and 126GiB memory.

Please reach out to any of the authors for access to the code. 

The checkpoints we used are available at \href{http://bit.ly/GAE-checkpoints}{\texttt{http://bit.ly/GAE-checkpoints}}.

\end{document}